\documentstyle[epsfig,amsmath,prl,aps]{revtex}

\begin{document}

\bibliographystyle{prsty}

\draft

\title{Instantaneous Normal Mode analysis of liquid HF}

\author{G. Garberoglio and R. Vallauri \cite{email}}
\address{Dipartimento di Fisica dell'Universit\`a di Trento and Istituto
Nazionale di Fisica della Materia \\ Via Sommarive 14, I-38050 Povo (TN),
Italy}

\date{January 19, 2000. Revised March 21, 2000}

\maketitle

\begin{abstract}
We present an Instantaneous Normal Modes analysis of liquid HF aimed to
clarify the origin of peculiar dynamical properties which are supposed to stem
from the arrangement of molecules in linear hydrogen-bonded network. The
present study shows that this approach is a unique tool for the understanding
of the spectral features revealed in the analysis of both single molecule and
collective quantities. For the system under investigation we demonstrate the
relevance of hydrogen-bonding ``stretching'' and fast librational motion
in the interpretation of these features.
\end{abstract}


In recent years the analysis of Instantaneous Normal Modes (INM) of normal and
supercooled liquids has given a sound improvement to the understanding of the
microscopic processes underlying the dynamical properties of these
systems. Applications of the method are to be found in the calculation of
macroscopic quantities through the knowledge of the density of states
(e.g. the diffusion coefficient \cite{keyes_diff}) and in the interpretation
of atomic motion through the inspection of the eigenvectors
\cite{madden,sciortino,mazzacurati}.
This last method appears to be a unique tool for the interpretation of
dynamical features (single molecule or collective) of a disordered system in
terms of correlated motions of its constituents. In particular it is
interesting to explore how the presence of locally ordered units is reflected
in the time behaviour of, for example, the velocity autocorrelation function
(VACF). Attempts in this direction have to be found in the analyses of:
$i)$  
the correlation function of the projection of the centre of mass (CoM)
velocity of water molecules along the directions of the normal coordinates of a
cluster of three molecules \cite{sv};
$ii)$
the projection of INM eigenvectors onto the totally symmetric displacement
coordinates of ZnCl$_4^{2-}$ tetrahedral units \cite{madden}. 
Moreover the INM approach has been exploited to describe the dynamical
features of molecular liquids (e.g. diatomic Lennard--Jones \cite{buchner},
CS${}_2$ 
\cite{keyes_cs2}) including hydrogen bonded systems like water
\cite{water_inm,sciortino}. On the other hand recent molecular dynamics (MD)
simulations of HF have revealed peculiar dynamical features, e.g. a peak at
$\simeq$ 50 ps${}^{-1}$
in the spectra of both collective and single molecule correlation
functions \cite{bstv,bbstv}. In the collective longitudinal and transverse
current spectra this mode appears to have an optical-like character, since its
frequency is found to be independent of the wavevector. For the interpretation
of the CoM VACF spectrum this peak has been
related to the relative motion of two nearest neighbour molecules
\cite{bbstv}. Since nearest neighbours are also hydrogen bonded it is tempting
to assign this dynamical feature to hydrogen bonding ``stretching''. As a
matter of fact HF molecules have been demonstrated to form irregular zig-zag
chains of different size, being this peculiar clustering favoured by the
geometry of the molecule and the strong electrostatic interaction
\cite{jv_nonpol}. A dynamical characterisation of these ordered units has not
yet been given, so that any assignment of the spectral features to particular
dynamical process remains speculative.

In the present Letter we report the results of an INM analysis of liquid HF
aimed to give an answer to the above questions. It will be shown to which
extent the presence of irregular chains is reflected in the INM spectra and
how the appearance of optical-like modes can be understood from the analysis
of the short time dynamics naturally expressed by the INM eigenvectors.

The use of the INM analysis starts from the solution of the eigenvalue
problem
\begin{equation}
\omega^2 {\sf T} {\bf e} = {\sf K} {\bf e}
\label{eq:eigenproblem}
\end{equation}
where ${\sf T}$ is the mass matrix of the system and ${\sf K}$ is the matrix
of the second derivatives of the potential energy. Here ${\bf e}$ represents a
multidimensional vector which specifies the instantaneous configuration of the
system, i.e. in general the three CoM coordinates and Euler angles of the
assumed rigid molecule. An average over many independent configurations has to
be performed in order to obtain the distribution of eigenfrequencies and any
dynamical quantity derived from the knowledge of the eigenvector ${\bf e}$. In
order to avoid spurious effects originating in the fact that the mass matrix
${\sf T}$, which depends on the sine of the polar angle, is not guaranteed to
be strictly positive definite at each time, we have used the orientational
coordinates described in Ref. \cite{keyes_irc} to evaluate the derivatives of
the potential energy.

We have calculated the INM for liquid HF (at T = 203 K and $\rho$ = 1.178
g/cm${}^3$), using 60 configurations separated by 2 ps with N = 108
molecules. The potential model, reported in
Ref. \cite{jv_hfstruct}, accounts for the interaction of three fractional
charges plus a Lennard--Jones contribution and it has been shown to
satisfactorily reproduce the thermodynamics and structure of the liquid.
Due to the complexity of the interaction potential we have evaluated the
matrix of the second derivatives numerically by displacing each generalized
coordinate with an increment $h$ = 10${}^{-5}$ \AA\ since it gives a good
stability of the matrix elements and no unusually high eigenvalues. We also
checked the accuracy of the results by the presence of three eigenmodes with
zero frequency. Our value of the increment is in accordance with the one used,
for example, in Ref. \cite{keyes_tos}. The generalized eigenvalue problem was
resolved using standard methods \cite{nr}.

\begin{figure}
\centerline{
\epsfig{file=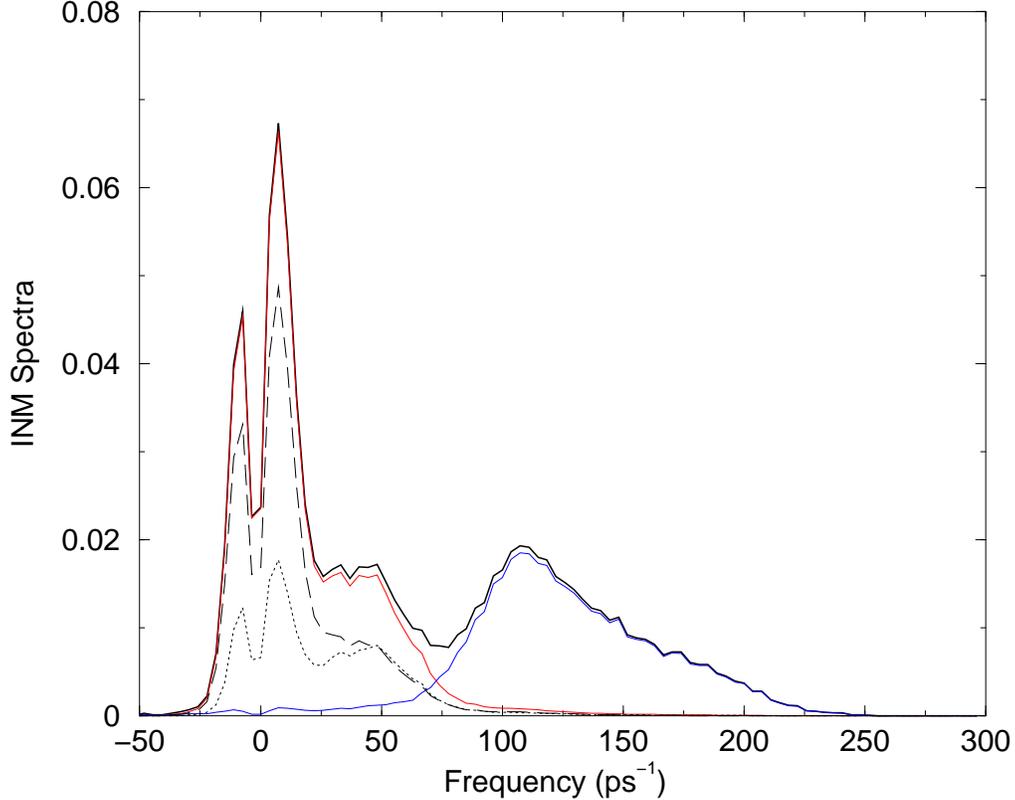,width=0.75\linewidth}
}
\caption{Spectrum of the INM, with imaginary frequencies reported in the
negative part, as usual. The thick line is the total spectrum, and has
been divided in translational and rotational components, the latter of which
appears only at high frequency. The dotted and dashed lines correspond to the
contribution parallel and perpendicular to the molecular axis respectively.}
\label{fig:inm_spectrum}
\end{figure}

The INM spectrum is presented in Fig. \ref{fig:inm_spectrum}. A comparison
with the results for water at room temperature (reported in
Ref. \cite{water_inm}) points out some important features. The contribution of
imaginary modes (conventionally reported on the negative axis) is
substantially larger in HF than in H${}_2$O (16\% against 6\%), a result
consistent with the relatively higher diffusion coefficient found in HF
\cite{bbstv}. Moreover there is a much more definite separation between
translational and rotational contributions defined by:
\begin{eqnarray}
\rho_T(\omega) & = & \left\langle 
\sum_i \sum_{\mu = 1,2,3} ({\bf e}^\alpha_{i \mu})^2
\delta(\omega - \omega_\alpha) \right\rangle\\
\rho_R(\omega) & = & \left\langle \sum_i \sum_{\mu = 4,5} ({\bf e}^\alpha_{i
\mu})^2 \delta(\omega - \omega_\alpha) \right\rangle
\end{eqnarray}
where ${\bf e}^\alpha_{i \mu}$ is the component of the eigenvector
corresponding to the frequency $\alpha$, referred to the $i$-th molecule and
to coordinate component $\mu$ ($\mu$ = 1,2,3 CoM, $\mu$ = 4,5
rotational coordinates).

The rotational spectrum extends from 50 to 250 ps${}^{-1}$, whereas in water
it goes from 0 to 180 ps${}^{-1}$. The translational spectrum of HF shows a
clear second maximum at $\simeq$ 50 ps${}^{-1}$ absent in the spectrum of
water where one can only notice an asymmetry of the low frequency maximum with
a larger content at higher frequency. We believe (and demonstrate in the
later) that this secondary maximum is related to the ``stretching'' of
hydrogen bond between two HF molecules. A comparison with the spectrum of the
CoM VACF \cite{bbstv} points out that the peak at $\simeq$ 50
ps${}^{-1}$ is much less pronounced and separated in the translational
component of the INM.


\begin{figure}
\centerline{
\epsfig{file=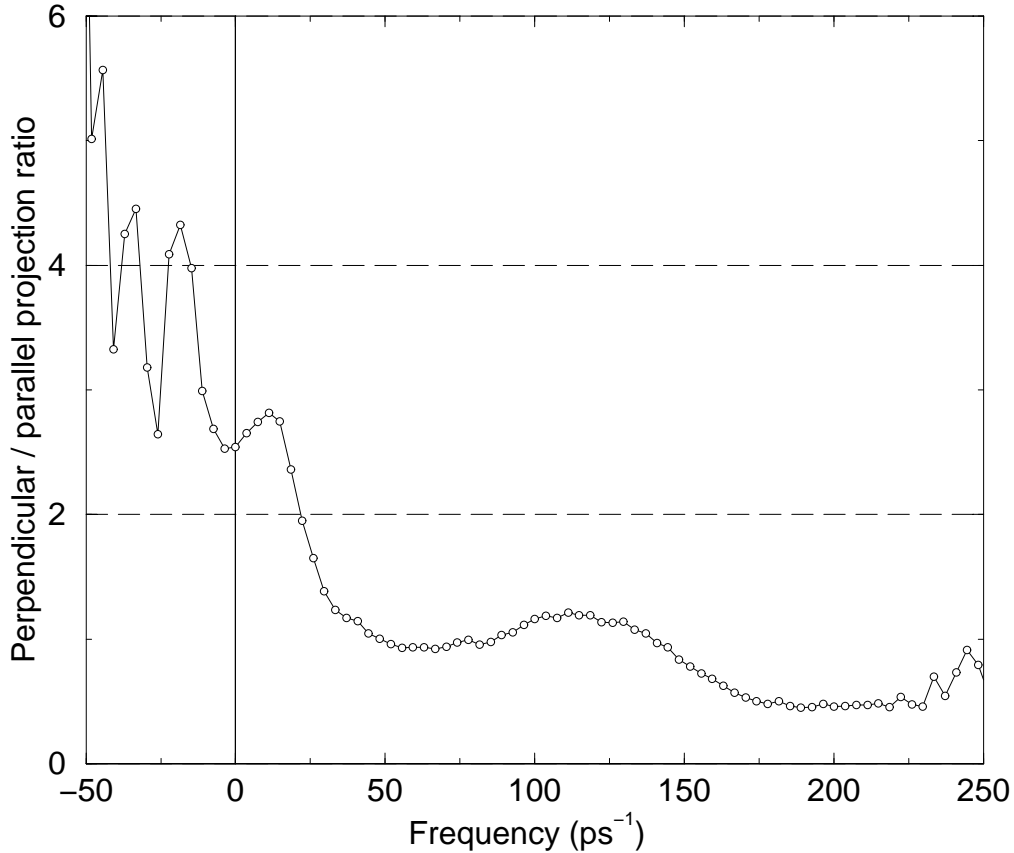,width=0.75\linewidth}
}
\caption{The ratio between $\rho_\perp(\omega)$ and
$\rho_\parallel(\omega)$ (see Eq. \ref{eq:parproj}). It should be 2 for an 
isotropic system.}
\label{fig:par_perp}
\end{figure}

The strong anisotropy of the single molecule dynamics is shown by performing a
projection of the translational component of the eigenvector along the
directions parallel and perpendicular to the molecular axis:
\begin{equation}
\rho_\parallel(\omega)  =  \left\langle 
\sum_i \sum_{\mu = 1,2,3} ({\bf e}^\alpha_{i \mu} {\bf u}_{i \mu})^2
\delta(\omega - \omega_\alpha) \right\rangle
\label{eq:parproj}
\end{equation}
where ${\bf u}_i$ is a unit vector along the symmetry axis of molecule $i$
and we have also defined $\rho_\perp(\omega) = \rho_T(\omega) -
\rho_\parallel(\omega)$.
If the molecular motion were isotropic, the ratio
between the perpendicular and parallel contributions would be equal to
two. Fig. \ref{fig:par_perp} shows that below 20 ps${}^{-1}$ this ratio
becomes larger than two going over four for modes at negative frequencies:
a result which points out that molecules can diffuse more freely in the
direction perpendicular to their axis rather than in the parallel one. Beyond
50 ps${}^{-1}$ the ratio becomes equal or less than one, thus revealing that
the motion of the molecules in this frequency range occurs, on a large extent,
in the direction of the molecular axis. Such an observation will be relevant
in discussing the arrangement of molecule along irregular chains as revealed
by the subsequent analysis of the INM eigenvectors. As a final remark we wish
to point out that the clear separation between translational and rotational
components can be considered as a print of hydrogen bonded systems. In fact it
is present in water and HF but not in a linear Lennard--Jones diatomic which
has been investigated in \cite{buchner}.

\begin{figure}
\centerline{
\epsfig{file=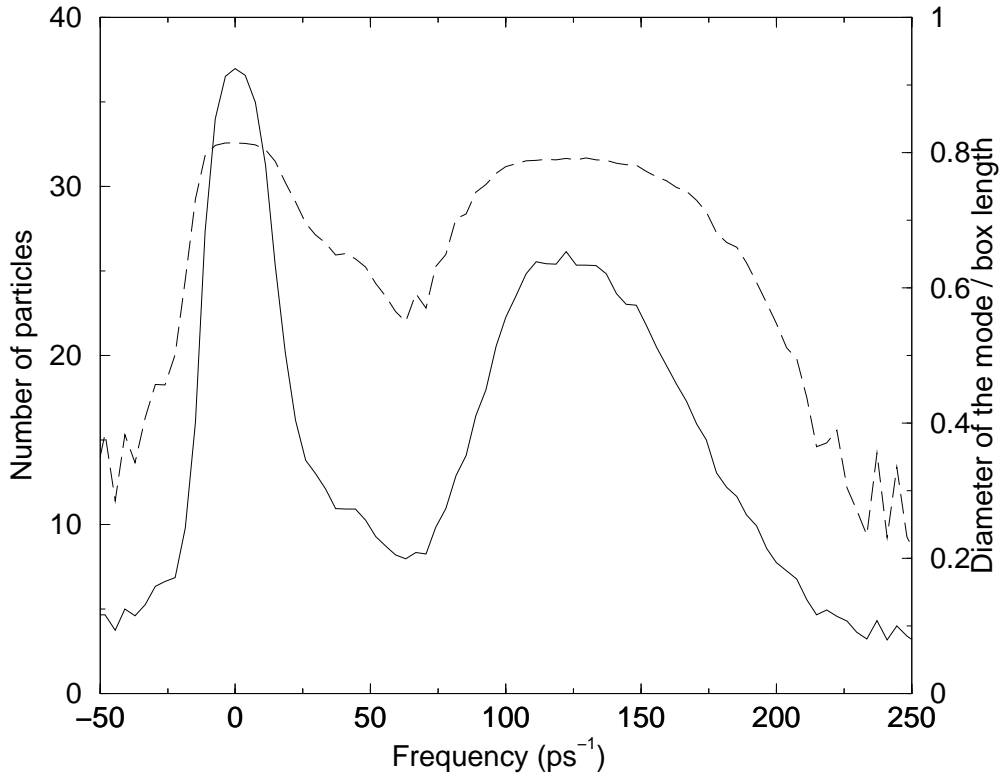,width=0.75\linewidth}
}
\caption{Solid line (left scale): Frequency distribution of the particles
participating in the modes. Dashed line (right scale): the distribution of the
mode radii, normalized with the simulation box length (14.504 \AA).}
\label{fig:part_n_radius}
\end{figure}

To analyse in more detail the structure of the INM we need to determine
which molecules participate in each given mode. We assume the following
criterion: the molecule $i$ said to belong to mode $\alpha$ if the condition
$
\sum_\mu ( {\bf e}_{i \mu}^\alpha )^2 > \frac{1}{N}
\label{eq:partmode}
$
is fulfilled. The distribution of the number of molecules per mode is reported
in Fig. \ref{fig:part_n_radius} (solid line) as a function of
frequency. Knowing the particles participating in a given mode, we can
determine the spatial localisation of the mode.


If ${\bf r}_i$ is the CoM position of particle $i$ and ${\cal
S}_\alpha$ is the set of particles involved in the mode we can define a
``radius'' for the mode as the maximum distance between two particles i.e.
$
R_\alpha = {\mathrm max}_{i,j \in {\cal S}_\alpha} | {\bf r}_i - {\bf r}_j |
\label{eq:mode_radius}
$ 
, the distribution of which is reported in Fig. \ref{fig:part_n_radius}
(dashed line). A comparison of the behaviour of the two quantities
indicates a clear correspondence between participation ratio and extension
of the modes. A lower number of participating molecules is accompanied by a
smaller radius of the mode. Surprisingly enough, however, in the range of
$\simeq$ 50 ps${}^{-1}$ the modes has an extension of about half of the box
length even if the number of molecules participating is smaller than ten.

A better understanding of this localisation problem can be obtained by
performing a projection of a mode onto the hydrogen bonded chains present in
the system, where hydrogen bonding is defined by the same
energetic criterion adopted in Ref. \cite{jv_hfstruct}. If ${\cal C}_c$
denotes the set of molecules belonging to the chain $c$ we define the
projection of the mode $\alpha$ on the chain as
\begin{equation}
P^\alpha_{{\cal C}_c} = 
\sum_{i \in {\cal C}_c, \mu} ( {\bf e}^\alpha_{i \mu} )^2
\end{equation}
which is one if the mode $\alpha$ is localised on the chain $c$, and is zero
if the mode $\alpha$ involves molecules not belonging to the chain. Clearly if
we consider the maximum of this projection taken on the set of all the chains,
we can see whether the modes are localised on some chain or not. 

\begin{figure}
\centerline{ \epsfig{file=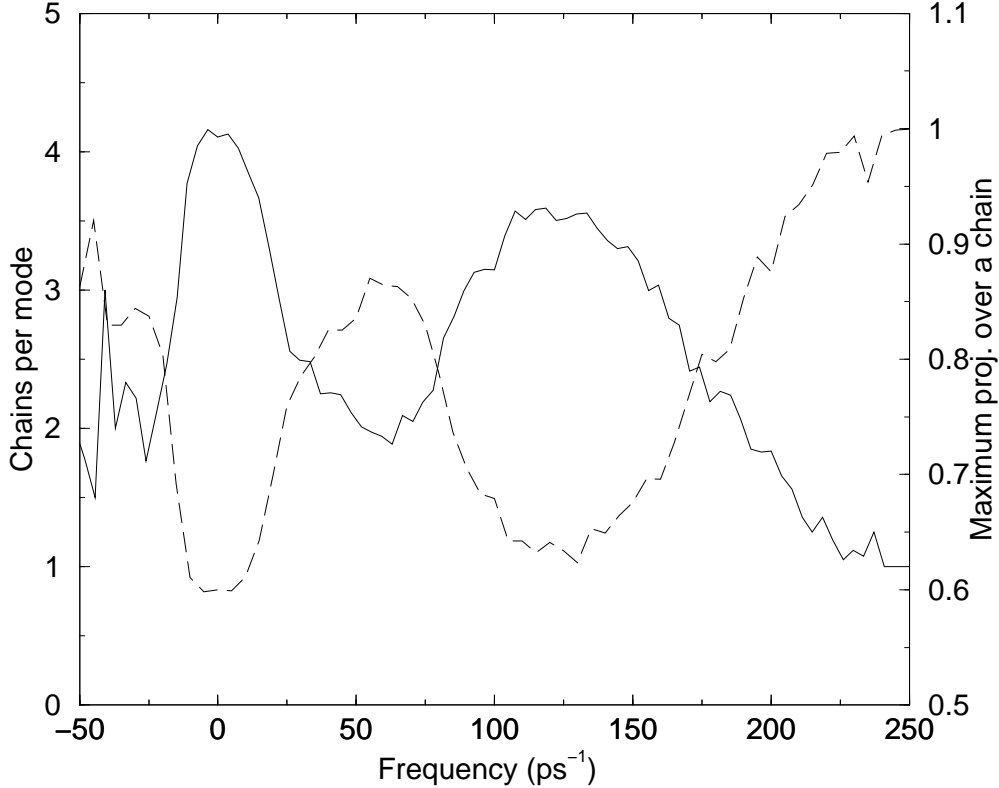,width=0.75\linewidth} }
\caption{Solid line (left scale): Frequency distribution for the number of
chains involved in the modes. Dashed line (right scale): Frequency dependence
of the mean value of the maximum projection of the eigenmodes onto a chain.}
\label{fig:maxproj_n_chain}
\end{figure}

The result presented in Fig. \ref{fig:maxproj_n_chain} (dashed line),
indicates that only in some frequency range the modes show a high degree of
localisation on a single chain, in particular in the same range where a
minimum partecipation ratio occurs and the radius of the mode is lower (see
Fig. \ref{fig:part_n_radius})


The idea that modes at particular frequencies are strongly correlated to the
presence of chains is confirmed by looking at the distribution of the number
of chains involved in a mode as reported in Fig. \ref{fig:maxproj_n_chain}
(solid line). In the range $\simeq$ 50 ps${}^{-1}$ (and in the region of
imaginary frequencies) we find the modes are spread over two distinct chains;
at the highest frequencies only one chain per mode is involved. Since in this
range only four molecules are participating to the mode (see
Fig. \ref{fig:part_n_radius}) we can conclude that these modes are confined on
the ring chains (tetramers) which are found to be particularly stable
\cite{jv_hfstruct}.

Having said that, it is also evident that most of the modes are far from being
localized over a single chain, e.g. all the modes whose maximum projection is
less than 0.80. This result is in accordance with other INM studies of network
forming systems \cite{madden}, where it is shown that the modes do not
typically reproduce the behaviour that could be expected on the basis of the
group properties of the network, but generally have some sort of mixed
character.

\begin{figure}
\centerline{
\epsfig{file=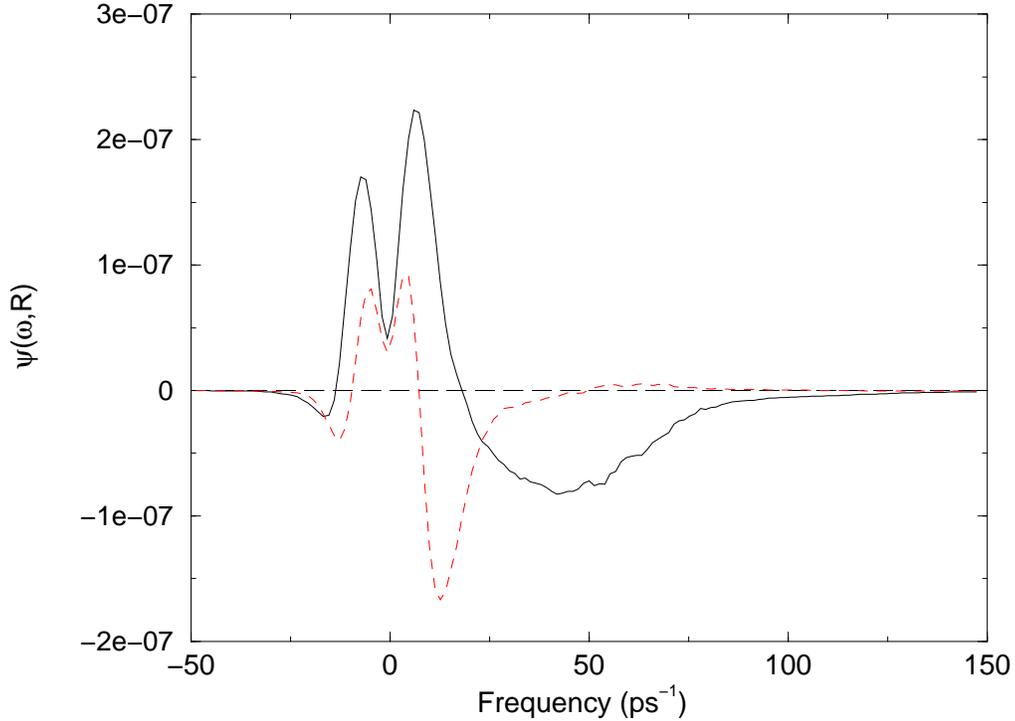,width=0.75\linewidth}
}
\caption{The value of the spatial correlation between particles in various
normal mode (see Eq. \ref{eq:spat_corr}). The solid line shows correlation
between the molecules being in the first shell, and the dashed line shows
correlation between molecules in the second shell (multiplied by a factor
of 5). The zero-frequency modes (corresponding to an overall translation
of the system) have been dropped from the calculation.}
\label{fig:psi}
\end{figure}

In order to characterize the spatial correlation of the molecular
displacements through the INM eigenvectors, we have examined the following quantity:
\begin{equation}
\psi(R,\omega) =
\left\langle
\frac{1}{3N} \sum_\alpha \frac{1}{n(R)} 
\sum_{|{\bf r}_i - {\bf r}_j| \in R} \sum_{\mu = 1,2,3}
{\bf e}_{i \mu}^\alpha {\bf e}_{j \mu}^\alpha
\delta(\omega - \omega_\alpha) 
\right\rangle
\label{eq:spat_corr}
\end{equation}
where $R$ denotes a spatial range (e.g. first or second shell of nearest
neighbours, defined through the minima of the CoM pair correlation) and $n(R)$
the number of pairs present in that range for a given configuration. This
quantity is a sort of mean value of the scalar product of the displacement of
the particles being nearest or next-to-nearest neighbours in a given normal
mode and at least at short time characterises in an exact way the motion of
the molecules participating to the particular mode. It is of course positive
if the particles move in phase and negative for out-of-phase
displacements. The results are shown in Fig. \ref{fig:psi}.


We notice that the nearest neighbours have opposite phases in a broad
frequency range around 50 ps${}^{-1}$, a behaviour in accordance with the
presence of the ``stretching'' mode similar to the one observed in the
solid \cite{hf_solid} and consistent with the fact that, in this range of
frequency, the molecules move preferentially along the direction of the
molecular axis. We notice that the next-to-nearest neighbours are not very
much correlated in this range, a signature of the fact that those modes are
somehow localised over a chain (see Fig. \ref{fig:maxproj_n_chain}) and do
not involve many molecules, in agreement with the value of the
participation ratio (see Fig. \ref{fig:part_n_radius}). The present result
unambiguously confirms the optical-like character of the mode at 50
ps${}^{-1}$ present in the longitudinal and transverse current spectra
reported in Ref. \cite{bstv}.

In a previous investigation of the collective properties of liquid HF
\cite{bstv} it has been shown that the longitudinal spectra have a peak,
$\omega_{max}$, at low frequency, which changes linearly with the
wavevector $k$. Its value is found to remain lower that 10 ps${}^{-1}$. The
corresponding phase velocity $v_{ph} = \omega_{max} / k$ turns out to be
somewhat higher than the ultrasonic (hydrodynamic) counterpart, but
compatible with the presence of a positive anomalous dispersion as in the
case of monatomic liquids (e.g. liquid metals). This feature is normally
interpreted in terms of overdamped acoustic modes propagating in the system
at wavevectors well beyond the hydrodynamic range. The results of the
present INM analysis are consistent with such an interpretation, since they
show that in the range below 10 ps${}^{-1}$: 1) the partecipation ratio is
large (see Fig. \ref{fig:part_n_radius}), 2) the short time displacements
of nearest and next-to-nearest neighbors are in phase as one would expect
from molecules participating to a collective (acoustic-like) motion (see
Fig. \ref{fig:psi}).

In conclusion the INM analysis of HF has revealed how the dynamical features
are affected by the presence of topological chains of hydrogen-bonded
molecules. In certain ranges of frequencies there is a strict correspondence
between INM modes and chains, where these modes are found to be
localised. This result has allowed to give a sound interpretation to a
feature present both in the VACF and collective currents, namely a
peak at 50 ps${}^{-1}$ in the corresponding spectra. Such a
characteristic can in fact be assigned to the ``stretching'' of hydrogen
bonding of first neighbouring molecules. This finding reveals the
``optical'' character of the collective mode present in the currents. Low
frequency modes are found to be spread over the whole system
involving several different chains, however they are
confined to frequencies not much higher than 10 ps${}^{-1}$ in agreement
with previous findings derived from the analysis of the longitudinal and
transverse currents.
The very high frequency (rotational) modes ($\simeq$ 200 ps${}^{-1}$)
are demonstrated to be confined over the highly stable tetramer chains.

Finally, we wish to stress the fact that our present study has the
potential to pave the way for an unambiguous interpretation of the
dynamical feature of other hydrogen bonded systems (e.g. water) which are
still a matter of a large debate.



\end{document}